\documentclass[a4paper,10pt]{article}
\usepackage{amssymb,amsmath,latexsym,array,graphicx}

\def\DIS{\displaystyle}

\usepackage{theorem}
\theoremstyle{break}
\newtheorem{Theorem}{Theorem}
\newtheorem{Proposition}{Proposition}
\newtheorem{Lemma}{Lemma}
\def\Proof{\hfil\break{\bf Proof}\quad}
\def\qed{\hfill\hbox{$\Box$}\vspace{10pt}\break}

\def\Z{{\mathbb Z}}

\begin{document}
\title{Correlation function for a periodic box--ball system }
\author{Jun Mada${}^1$ and Tetsuji Tokihiro${}^2$ \\ \\
${}^1$ College of Industrial Technology, \\
  Nihon University, 2-11-1 Shin-ei, Narashino, Chiba 275-8576, Japan \\
${}^2$ Graduate School of Mathematical Sciences, \\
  University of Tokyo, 3-8-1 Komaba, Tokyo 153-8914, Japan} 

\date{}

\maketitle

\begin{abstract}
We investigate correlation functions in a periodic box--ball system. 
For the two point functions of short distance, we give explicit formulae 
obtained by combinatorial methods. 
We give expressions for general $N$-point functions 
in terms of ultradiscrete theta functions. 
\end{abstract}

\section{Introduction}
\label{sec:introduction}
Quantum integrable systems such as quantum integrable spin chains 
and solvable lattice models are systems whose Hamiltonians 
or transfer matrices can be diagonalised and for which eigenstates 
or free energies can be explicitly obtained~\cite{Baxter}. 
To investigate physical properties of these systems, 
such as {\it e.g.} the linear response to external forces, however, 
we further need to evaluate correlation functions for these systems.
This is one of the main problems in the field of quantum integrable systems and in fact, 
obtaining correlation functions is even fairly difficult for the celebrated XXZ model 
or the 6 vertex model \cite{JM}. 

A periodic box-ball system (PBBS) is a soliton cellular automaton 
obtained by ultradiscretizing the KdV equation~\cite{TS, YT}. 
It can also be obtained at the $q \to 0$ limit 
of the generalized 6 vertex model~\cite{FOY, HHIKTT}. 
Hence, from the view point of quantum integrable lattice models, 
it is interesting and may actually give some new insights into the correlation functions 
of the vertex models themselves, to obtain correlation functions of the PBBS.
In this paper, we give expressions for $N$-point functions for the PBBS, 
using combinatorial methods and the solution for the PBBS 
expressed in terms of the ultradiscrete theta functions. 

\bigskip

\begin{figure}[t]
 \begin{center}
 \includegraphics[width=.6\linewidth]{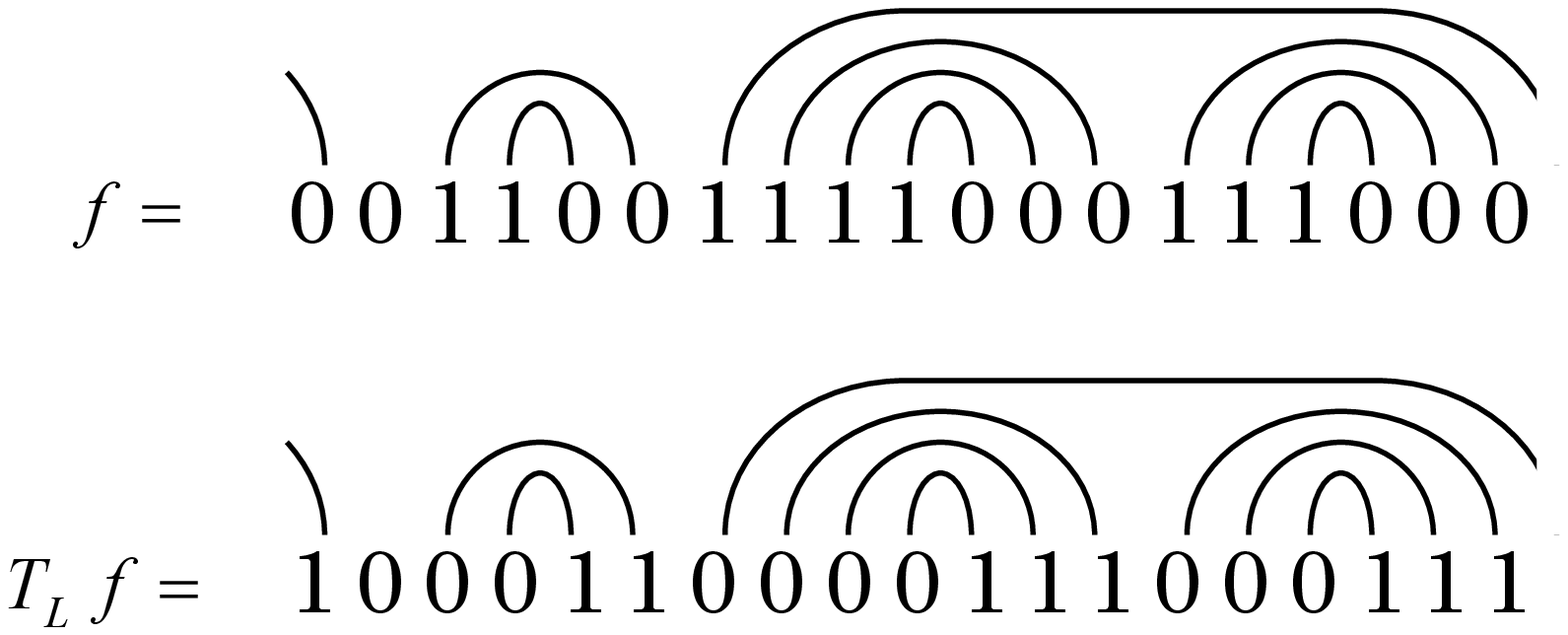} 
 \caption{Definition of $T_L$ for $f\in\Omega_L$}
 \label{fig:TL}
 \end{center}
\end{figure}

The PBBS can be defined in the following way. 
Let $L\ge3$ and let $\DIS \Omega_L=\big\{\,f\,|\,f:[L]\to\{0,1\}\ \mbox{such that}\ 
\sharp f^{-1}(\{1\})<L/2\,\big\}$ where $[L]=\{1,2,\ldots,L\}$. 
When $f\in \Omega_L$ is represented as a sequence of $0$s and $1$s, we write 
\[
 f(1)f(2)\ldots f(L). 
\]
The mapping $T_L:\Omega_L\to \Omega_L$ is defined as follows (see Fig.~\ref{fig:TL}): 
\begin{enumerate}
\item In the sequence $f$ find a pair of positions $n$ and $n+1$ 
      such that $f(n)=1$ and $f(n+1)=0$, and mark them;
      repeat the same procedure until all such pairs are marked.
      Note that we always use the convention that the position $n$ is defined in $[L]$, 
      {\it i.e.} $n+L\equiv n$. 
\item Skipping the marked positions we get a subsequence of $f$; 
      for this subsequence repeat the same process of marking positions, 
      so that we get another marked subsequence.
\item Repeat part 2 until one obtains a subsequence consisting only of $0$s. 
      A typical situation is depicted in Fig.~\ref{fig:TL}. 
      After these preparatory processes, 
      change all values at the marked positions simultaneously; 
      One thus obtains the sequence $T_Lf$. 
\end{enumerate}
Sometimes we shall write $T_L^tf$ for $\underbrace{T_L(\cdots(T_L(T_L}_{t}f)))$. 
The pair $(\Omega_L,T_L)$ is called a PBBS of length $L$~\cite{YT,MIT2006}. 
An element of $\Omega_L$ is called a state, and the mapping $T_L$ the time evolution. 

An $N$-point function of the PBBS with $M$ balls may be defined as follows. 
\begin{equation*}
 \langle s_1,s_2,\ldots,s_N \rangle:=\frac{1}{Z_H}\sum_{f\in\Omega_{L;M}}
 \mbox{e}^{\sum_{k=1}^L H_k(f)} f(s_1)f(s_2)\cdots f(s_N)
\label{1.Npoint}
\end{equation*}
where $\DIS \Omega_{L;M}:=\left\{ f \in \Omega_L\, \big|\, \sharp f^{-1}(\{1\})=M\,\right\} $, 
$Z_H:=\sum_{f \in \Omega_{L;M}}
\mbox{e}^{\sum_{k=1}^L H_k(f)}$ and $H_k(f)$ is the $k$th energy of the state $f$,
which is proportional to the number of $k$th arc lines 
defined when determining the time evolution rule~\cite{YT}, 
or the $k$th conserved quantity of the PBBS~\cite{TTS}. 
(Note that $H_k(f)$ is essentially equal to the energy function 
for the transfer matrix of the crystal lattice models with $k+1$ 
states on a vertical link~\cite{FOY, MIT2006}.) 
Noticing the fact that $\Omega_{L;M}=\bigsqcup_Y \Omega_Y$, 
\[
 \langle s_1,s_2,\ldots,s_N \rangle=\frac{1}{Z_H}\sum_{Y}\sum_{f\in\Omega_Y}
 \mbox{e}^{\sum_{k=1}^L H_k(f)} f(s_1)f(s_2)\cdots f(s_N),
\]
where $Y$ are partitions of $M$ corresponding to the conserved quantities of the PBBS. 
(See Section~\ref{sec:combinatorial}.)
Since, for $f_i \in \Omega_{Y_i}$ $(i=1,2)$, ${}^\forall k \ H_k(f_1)=H_k(f_2)$ 
$(k=1,2,3,\ldots)$ implies $Y_1=Y_2$ and vice versa, by choosing a state $f_Y$ in $\Omega_Y$
we can write
\[
\langle s_1,s_2,\ldots,s_N \rangle=\frac{1}{Z_H}\sum_{Y}\mbox{e}^{\sum_{k=1}^L H_k(f_Y)} 
\sum_{f\in\Omega_Y} f(s_1)f(s_2)\cdots f(s_N).
\]
Thus, to obtain correlation functions of PBBS, we have only to evaluate those on the set $\Omega_Y$:
\begin{equation}
\langle s_1,s_2,\ldots,s_N \rangle_Y
:=\frac{1}{|\Omega_Y|}\sum_{f\in\Omega_Y} f(s_1)f(s_2)\cdots f(s_N).
\label{1.NpointY}
\end{equation}
We also point out that if we put ${}^\forall k,\, {}^\forall f,\ H_k(f)=0$, 
$N$-point functions become trivial; 
\[
\langle s_1,s_2,\ldots,s_N \rangle = \frac{_{L-N}C_{M-N}}{_LC_M}
=\frac{M(M-1)\cdots(M-N+1)}{L(L-1)\cdots(L-N+1)}.
\]
In the following sections we shall evaluate \eqref{1.NpointY}.

\bigskip

First we summarize some useful properties of the PBBS.
We say that $f$ has (or that there is) a $10$-wall at position $n$ 
if $f(n-1)=1$ and $f(n)=0$. Let the number of the $10$-walls be $s$ 
and the positions be denoted by $a_1>a_2>\cdots>a_s$. 
Then, we have the following proposition: 

\begin{Proposition}[\cite{MIT2008}]
\label{prop:MIT2008}
\begin{eqnarray}
 (T_L^tf)(n) &= & \eta_{n+1}^{t-1}-\eta_{n+1}^t-\eta_n^{t-1}+\eta_n^t, 
 \nonumber \\[2mm]
 \eta_n^t &= & \max_{{m_i\in\Z}\atop {i\in[s]}} 
 \left[\ \sum_{i=1}^s m_i\big( b_i+tW_i-n \big) 
 -\sum_{i=1}^\ell \sum_{j=1}^\ell m_i\Xi_{ij}m_j\ \right], 
 \label{aeta} \\[2mm]
 b_i &= & a_i+\sum_{j=1}^{i-1} 2\min\{W_i,W_j\}+W_i+\frac{Z_i}{2}, 
 \label{aetab} \\[2mm]
 \Xi_{ij} &= & \frac{Z_i}{2}\delta_{ij} +\min\{W_i,W_j\}, \nonumber \\
 Z_i &= & L-\sum_{j=1}^s 2\min\{W_i,W_j\}, \nonumber 
\end{eqnarray}
where $W_i$ denotes the amplitude of the ``soliton'' corresponding to $a_i$ 
obtained by the procedure explained in \cite{MIT2008}. 
\end{Proposition}
The set $\big\{W_i\big\}_{i=1}^s$ consists of quantities of the PBBS 
and $\eta_n^t$ is the ultradiscrete theta function~\cite{MZ}. 
We shall use Proposition~\ref{prop:MIT2008} for determining $N$-point functions 
in Section~\ref{sec:solution}. 

Next we introduce two procedures which are important in this paper. 
For a given $f\in\Omega_L$, a state $Ef=E(f)$ is defined to be
\[
(Ef)(n)=\left\{ \begin{array}{ll}
 \left\{ \begin{array}{ll}
 f(n) & (1\le n\le a_s-2), \\
 f(n+2k) & \left( \begin{array}{r} a_{s-k+1}-2k+1\le n\le a_{s-k}-2k-2 \\ 
                                   (k=1,2,\ldots,s-1) \\ \end{array} \right), \\
 f(n+2s) & (a_1-2s\le n\le L-2s), \\
\end{array} \right. & (a_s>1) \\ \\
 \left\{ \begin{array}{ll}
 f(n+1) & (1\le n\le a_{s-1}-3), \\
 f(n+2k+1) & \left( \begin{array}{r} a_{s-k}-2k\le n\le a_{s-k-1}-2k-3 \\ 
                                     (k=1,2,\ldots,s-2) \\ \end{array} \right), \\
 f(n+2s-1) & (a_1-2s+1\le n\le L-2s). \\
\end{array} \right. & (a_s=1) \\
\end{array} \right.
\]
The mapping $E:\Omega_L\to\Omega_{L-2s}$ is called the $10$-elimination. 
$Ef$ is a subsequence of $f$ obtained by eliminating all $10$-walls 
in $f$ simultaneously. 
For example, 
\begin{eqnarray*}
 f &= & 001111\underline{10}0001111\underline{10}0000000
 111\underline{10}01\underline{10}0001111
 \underline{10}011\underline{10}000111\underline{10}000000000, \\
 Ef &= & 001111~~~0001111~~~0000000111~~~01~~~0001111~~~011~~~000111~~~000000000 \\
 &= & 00111100011110000000111010001111011000111000000000.
\end{eqnarray*}

Its inverse process is called the $10$-insertion, 
$I(j_1,j_2,\ldots,j_d)=I_2\circ I_1(j_1,j_2,\ldots,j_d): 
\Omega_L\to \Omega_{L+2(d+s)}$ where $s$ is the number of $10$-walls in $f\in\Omega_L$. 
The $10$-insertion is defined as follows: 
Shifting the origin if necessary, we can assume that $f(L)=0$. 
For $\{j_1,j_2,\ldots,j_d\}\ (1<j_1<j_2<\cdots<j_d\le L+d)$, 
the mapping $I_1(j_1,j_2,\ldots,j_d):\Omega_L\to\Omega_{L+2d}$ is defined as 
\begin{align*}
 & (I_1(j_1,j_2,\ldots,j_d)f)(n) \\
 &\qquad =\left\{ \begin{array}{ll}
  1 & (n=L+2d-j_k-k+1), \\[1mm]
  0 & (n=L+2d-j_k-k+2), \\ [1mm]
  f(n) & (1\le n\le L+d-j_d), \\[1mm]
  f(n-2(d-k+1)) & (L+2d-j_k-k+3\le n\le L+2d-j_{k-1}-k+1), \\[1mm]
  f(n-2d) & (L+2d-j_1+2\le n\le L+2d) \\[1mm]
 \end{array} \right. 
\end{align*}
where $k\in [d]$; furthermore, $I_2:\Omega_{L+2d}\to\Omega_{L+2(d+s)}$ is defined to be 
\[
 (I_2f')(n) =\left\{ \begin{array}{ll}
  1 & (n=g_k+2(s-k)+1), \\[1mm]
  0 & (n=g_k+2(s-k)+2), \\ [1mm]
  f'(n) & (1\le n\le g_{s}), \\[1mm]
  f'(n-2(s-k+1)) & (g_k+2(s-k)+3\le n\le g_{k-1}+2(s-k)+2), \\[1mm]
  f'(n-2s) & (g_1+2s-2\le n\le L+2(d+s)) \\[1mm]
 \end{array} \right. 
\]
where $k \in [s],\ f'\equiv I_1(j_1,j_2,\ldots,j_d)f\in\Omega_{L+2d}$ and
\begin{align}
 g_k' &= \max \Big\{\,m\in[L+d]\,\Big|\,
            m=a_k-1+\sharp\big\{\,r\in[d]\,\big|\,L+d-j_r+1<m\,\big\}\,\Big\}, 
 \nonumber \\
 g_k &= g_k'+\sharp\big\{\,r\in[d]\,\big|\,L+d-j_r+1<g_k'\,\big\}. 
 \label{gk}
\end{align}
For example, 
\begin{eqnarray*}
 f &= & 0011100111000001101000111000000,\\
 I_1(3,11,25)f &= & 001110011\ast1000001101000\ast1110000\ast00 \\
 &= & 001110011\underline{10}1000001101000
 \underline{10}1110000\underline{10}00, \\
 I(3,11,25)f &= & 00111\fbox{\hskip-1.2mm 10\hskip-1.1mm}0011
 \underline{10}1\fbox{\hskip-1.2mm 10\hskip-1.1mm}0000011
 \fbox{\hskip-1.2mm 10\hskip-1.1mm}01\fbox{\hskip-1.2mm 10\hskip-1.1mm}000
 \underline{10}111\fbox{\hskip-1.2mm 10\hskip-1.1mm}0000\underline{10}00 \\
\end{eqnarray*}
where $\underline{10}$ and $\fbox{\hskip-1.2mm 10\hskip-1.1mm}$ denote 
the inserted $10$ at $f\mapsto I_1(j_1,j_2,\ldots,j_d)f$ 
and $I_1(j_1,j_2,\ldots,j_d)f\mapsto I_2(I_1(j_1,j_2,\ldots,j_d)f)$ respectively.

\section{One and two point functions obtained by combinatorial methods}
\label{sec:combinatorial}
We assume that $Y$ denoting the conserved quantities of $f\in\Omega_Y$, is the partition 
\[
\big(\underbrace{P_1,P_1,\cdots,P_1}_{n_1},\underbrace{P_2,P_2,\cdots,P_2}_{n_2},
\ldots\underbrace{P_\ell,P_\ell,\cdots,P_\ell}_{n_\ell}\big)
\]
where $P_1>P_2>\cdots >P_\ell\ge1$. 
Note that $Y$ is a partition of $M$, {\it i.e.} $M=\sum_{i=1}^\ell n_iP_i$. 
As mentioned in Section~\ref{sec:introduction}, 
we consider $N$-point functions \eqref{1.NpointY} of the PBBS,
\[
\langle s_1,s_2,\ldots,s_N \rangle_Y 
=\frac{1}{|\Omega_Y|} \sum_{f\in \Omega_Y} f(s_1)f(s_2)\cdots f(s_N).
\]
The value of $|\Omega_Y|$ is already known: 

\begin{Proposition}[\cite{YYT}]
\[
\big|\Omega_Y\big|=\frac{L}{L_0}
 \left( \begin{array}{c} L_0+n_1-1 \\ n_1 \\ \end{array} \right)
 \left( \begin{array}{c} L_1+n_2-1 \\ n_2 \\ \end{array} \right)\cdots
 \left( \begin{array}{c} L_{\ell-1}+n_\ell-1 \\ n_\ell \\ \end{array} \right)
\]
where $L_0=L-2M,\ L_i=L_0+\sum_{j=1}^i 2n_j(P_j -P_{i+1})$ and $P_{\ell+1}=0$. 
\end{Proposition}

Since the $N$-point function $\langle s_1,s_{1}+d_1,\ldots,s_{1}+d_{N-1} \rangle_Y$ does not 
depend on the specific site $s_1$ 
(because of translational symmetry), we denote 
\[
 C_Y(d_1,d_2,\ldots,d_{N-1})\equiv 
 \langle s_1,s_{1}+d_1,\ldots,s_{1}+d_{N-1} \rangle_Y
\]
where $1\le d_1<d_2<\cdots<d_{N-1}< L$. 
Note that $C_Y(\emptyset)$ denotes the $1$-point function $\langle s_1 \rangle_Y$. 

\bigskip

\begin{Proposition}
$$C_Y(\emptyset)=\frac{M}{L}.$$
\end{Proposition}
\Proof 
Since $\sum_{n=1}^L f(n) =M$, 
\[
LC_Y(\emptyset) =\sum_{s_1=1}^L \langle s_1 \rangle_Y 
=\frac{1}{|\Omega_Y|} \sum_{f\in \Omega_Y} \sum_{n=1}^L f(n)
=\frac{1}{|\Omega_Y|} |\Omega_Y|M =M.
\]
\qed

Next we consider the $2$-point functions. 

\begin{Proposition}
$$C_Y(1) =\frac{M-s}{L}$$
where $s=\sum_{i=1}^\ell n_i$. 
\end{Proposition}
\Proof 
Since $\sum_{n=1}^L f(n)f(n+1)=M-s$, 
$$LC_Y(1) =\frac{1}{|\Omega_Y|} \sum_{f\in \Omega_Y} \sum_{n=1}^L f(n)f(n+1) =M-s.$$
\qed

In order to investigate $C_Y(2)$, let us put
\begin{align*}
 k_i &:= \left\{ \begin{array}{ll} 
 n_j & (i=P_j), \\[1mm]
 0 & \mbox{otherwise,} \\
 \end{array} \right. \\
 \hat{k}_i &:= \sum_{j=i}^{P_1} k_i, \\
 \tilde{L} &:= L-2\hat{k}_1\quad (=L-2s),\\
 N_Y(2) &:= \sum_{i=3}^{P_1} k_i(i-2). 
\end{align*}
We also define 
\begin{align*}
V_{f_0} &:= \left\{\,f\in\Omega_Y\,\big|\,Ef=f_0\,\right\},\\[2mm]
G_2(f) &:= \sharp\left\{\,n\in[L]\,\big|\,f(n)f(n+2)=1\,\right\}. 
\end{align*}
The following lemma is the key to evaluating $C_Y(2)$.
\begin{Lemma}
\label{Lemma1}
Let
\[
V_{f_0}^{(j)}:=\left\{\, f \in V_{f_0} \, \big| \, G_2(f)=N_Y(2)+j \, \right\}.
\]
Then, if $V_{f_0}\ne\phi$, $V_{f_0}=\bigsqcup_{j=0}^{k_1} V_{f_0}^{(j)}$ 
and
\begin{equation}
\left| V_{f_0}^{(k_1-j)} \right|=\frac{\nu_j}{k_1!}
\label{2.numberofstates}
\end{equation}
where
\begin{eqnarray*}
 \nu_j &:= & \Bigg( \prod_{i=0}^{j-1} (\tilde{L}-2\hat{k}_2-i) \Bigg) 
   \Bigg( \prod_{i=0}^{k_1+j-1} (2\hat{k}_2+i) \Bigg) \\
 && \times \Bigg( \sum_{{0\le i_1<\cdots <i_j <k_1+j-1} \atop{i_h+1<i_{h+1}}} 
 \prod_{h=1}^j \frac{1}{(2\hat{k}_2+i_h)(2\hat{k}_2+i_h+1)} \Bigg). 
\end{eqnarray*}
\end{Lemma}
\Proof
When $f \in V_{f_0}$, there exists a set of positive numbers 
$\{j_i\}_{i=1}^{k_1}\,$ $(1<j_1<j_2<\ldots <j_{k_1}\le \tilde{L}+k_1)$ such that
\[
f=I(j_1,j_2,\ldots,j_{k_1})f_0.
\]
By examining the positions of $101$ and $111$, we find that
\[
G_2(f)=N_Y(2)+\gamma+J
\]
where $\DIS \gamma=\gamma\big(f_0;\{j_i\}_{i=1}^{k_1}\big)$ is the number of \underline{$10$}s 
inserted into the positions adjacent to consecutive 1s, 
and $J=\sharp\big\{\,i\in[d-1]\,\big|\,j_i+1=j_{i+1}\,\big\}$. 
(See the table below.)
For example,
\[
f_0=001110000100110000
\]
and $f=I(5,6,14,15,18)f_0$, then
\begin{align*}
f&=0011110100010100011000111010100000 \\
 (&=00111\fbox{\hskip-1.2mm 10\hskip-1.1mm}\underline{10}00
    \underline{10\hskip-0.3mm}\hskip0.6mm\underline{\hskip-0.3mm10}
    001\fbox{\hskip-1.2mm 10\hskip-1.1mm}00
    11\fbox{\hskip-1.2mm 10\hskip-1.1mm}
    \underline{10\hskip-0.3mm}\hskip0.6mm\underline{\hskip-0.3mm10}0000\ ).
\end{align*}
In this example, $k_1=5, \ \hat{k}_2=3,\ N_Y(2)=3$, $\gamma=2$ and $J=2$.
Since $0 \le \gamma+J \le k_1$, we have the decomposition 
$V_{f_0}=\bigsqcup_{j=0}^{k_1} V_{f_0}^{(j)}$.
\begin{equation*}
 \begin{array}{|c|c|c|c|c|} \hline & \multicolumn{4}{c|}{} \\[-3mm]
 f_0 & \multicolumn{4}{c|}{00111000} \\[1mm] \hline & \multicolumn{4}{c|}{} \\[-3mm]
 G_2(f_0) & \multicolumn{4}{c|}{1} \\ \hline
 \hline & & & & \\[-3mm]
 f=I(k)f_0 & 00\underline{10}111\fbox{\hskip-1.2mm 10\hskip-1.1mm}000 
           & 001\underline{10}11\fbox{\hskip-1.2mm 10\hskip-1.1mm}000 
           & 00111\fbox{\hskip-1.2mm 10\hskip-1.1mm}\underline{10}000 
           & 00111\fbox{\hskip-1.2mm 10\hskip-1.1mm}00\underline{10}0 \\[1mm] 
 {} & (k=7) & (k=6) & (k=4) & (k=2) \\[1mm] 
 \hline & & & & \\[-3mm] 
 G_2(f) & 3 & 2 & 3 & 2 \\
 {} & (\gamma=1,\,J=0) & (\gamma=0,\,J=0) & (\gamma=1,\,J=0) & (\gamma=0,\,J=0) \\ \hline 
 \end{array} 
\end{equation*}
To know $\DIS \left| V_{f_0}^{(j)} \right|$, 
we have only to count the number of states with $\gamma+J=j$.

For $k_1=1$, $\DIS \big| V_{f_0} \big|=\tilde{L}$.
Since there are $\hat{k}_2$ sets of consecutive 1s, 
$2 \hat{k}_2 $ states have $ \gamma+J=1$ $\ (\gamma=1,\ J=0)$
and the other $\tilde{L}-2\hat{k}_2$ states have $\gamma+J=0$ $\ (\gamma=0,\ J=0)$.

For $k_1=2$, let $f=I(j_1,j_2)f_0$.
As was seen in case $k_1=1$, there are $2 \hat{k}_2$ positions at which $\gamma+J$ 
can be increased by one.
If one $10$ pair is inserted in one of these positions, 
then there are $2\hat{k}_2+1$ positions for the other pair to increase $\gamma + J$ by one, 
and $\tilde{L}-2\hat{k}_2$ positions not to increase it.
On the other hand, if one $10$ pair is inserted at one 
of the $\tilde{L}-2\hat{k}_2$ non-increasing positions,
then there are $2\hat{k}_2 +2 $ positions for the other pair 
to increase $\gamma+J$ by one, and $\tilde{L}-2\hat{k}_2-1$ positions not to increase it. 
Hence, considering duplication of insertion, 
there are $\DIS (2\hat{k}_2)(2\hat{k_2}+1)/2!$ states with $\gamma+J=2$,
$\DIS \left[ (2\hat{k}_2)(\tilde{L}-2\hat{k}_2) 
+(\tilde{L}-2\hat{k}_2)(2\hat{k}_2 +2) \right]/2!$ states
with $\gamma+J=1$, and $\DIS (\tilde{L}-2\hat{k}_2)(\tilde{L}-2\hat{k}_2-1)/2!$ states 
with $\gamma+J=0$.

\begin{figure}[t]
 \begin{center}
 \includegraphics[width=.8\linewidth]{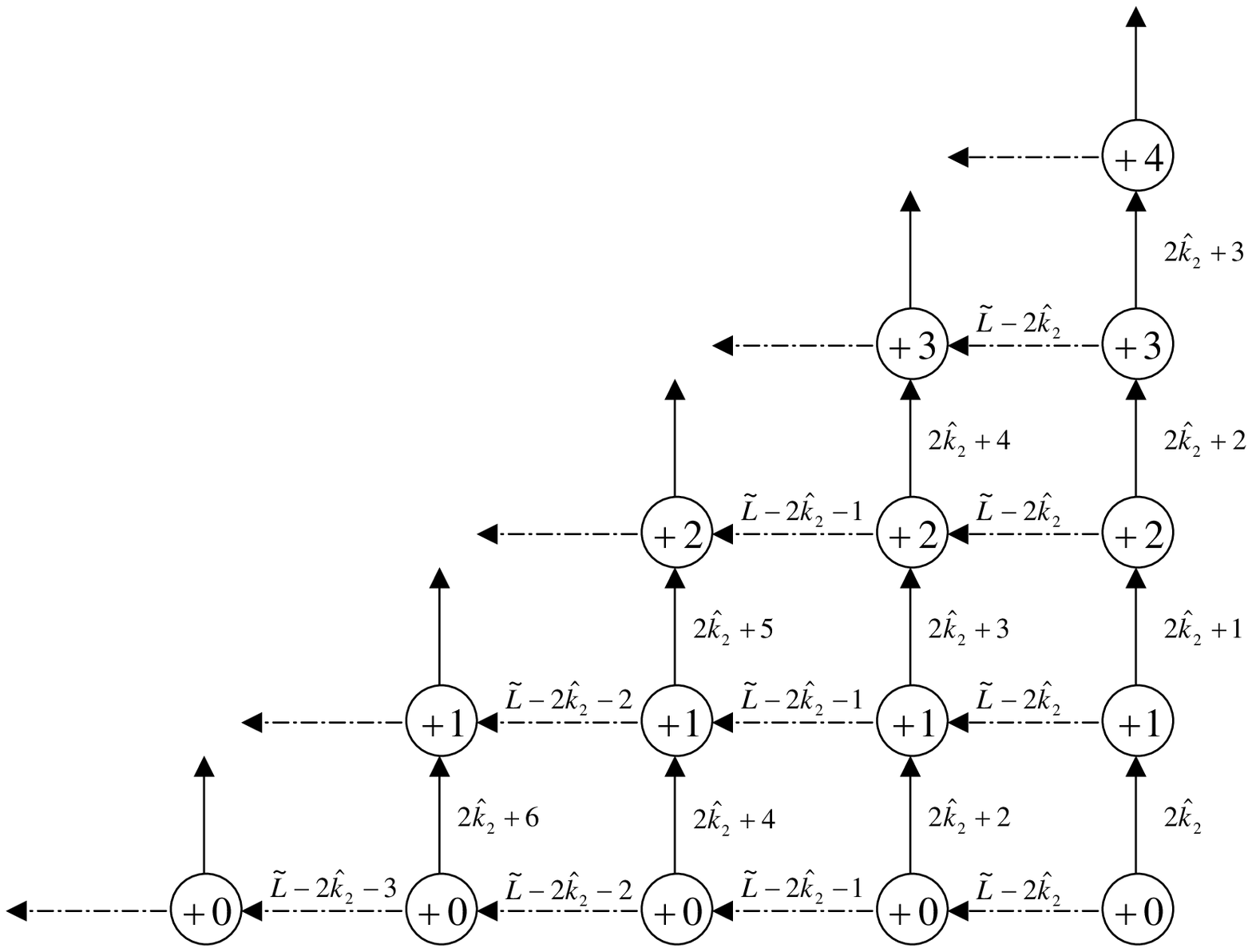} 
 \caption{A chart corresponding to $\gamma+J$ in the proof of Lemma~\ref{Lemma1}.}
 \label{fig:chart}
 \end{center}
\end{figure}

In general, we can proceed in a similar manner and, 
referring to the chart in Fig.~\ref{fig:chart}, we obtain \eqref{2.numberofstates}.
\qed

\begin{Proposition}
\begin{equation*}
 C_Y(2) 
 =\frac{\DIS \sum_{j=0}^{k_1} \nu_j\left( \sum_{i=3}^{P_1} k_i(i-2)+(k_1-j) \right)}
        {\DIS L\sum_{j=0}^{k_1} \nu_j}.
\label{cy(2)(2)}
\end{equation*}
\end{Proposition}
\Proof
From Lemma~\ref{Lemma1}, we see that if $V_{f_0}\ne\phi$, 
\[
\sum_{f \in V_{f_0}} \sum_{n=1}^L f(n)f(n+2)
=\sum_{j=0}^{k_1} \frac{\nu_j}{k_1!} \Big( N_Y(2)+(k_1-j) \Big)
\]
and 
\[
\big| V_{f_0} \big|= \sum_{j=0}^{k_1} \frac{\nu_j}{k_1!}.
\]
Since the right hand side of the last equation does not depend on $f_0$, 
and since any state $f \in \Omega_Y$ belongs to some $V_{f_0}$,
we obtain 
\[
LC_Y(2) =\frac{1}{|\Omega_Y|} \sum_{f\in \Omega_Y} \sum_{n=1}^L f(n)f(n+2) 
=\frac{\DIS \sum_{j=0}^{k_1} \nu_j\Big( N_Y(2)+(k_1-j) \Big)}
       {\DIS \sum_{j=0}^{k_1} \nu_j}.
\]
\qed

For $C_Y(d)\enskip (d\ge 3)\,$ we can use similar arguments based on elementary combinatorics. 
However, the expressions become more and more complicated when the difference $d$ increases. 
Instead in the next section we shall use Proposition~\ref{prop:MIT2008} to obtain expressions 
for general $N$-point functions.

\section{$N$-point correlation functions for the PBBS}
\label{sec:solution}
Let the state $f_0$ and the set ${\cal X}_Y\subset 
\Z_+^{n_1}\times \Z_+^{n_2}\times \cdots \times\Z_+^{n_\ell}\ (=\Z_+^s)$ be
\[
f_0=\underbrace{000\cdots00}_{L_0},
\]
and 
\begin{equation}
{\cal X}_Y:=\left\{ \big\{ x_i(k) \big\}_{i=1,k=1}^{\ell,\quad n_i}\,\left|
 \,\begin{array}{r} 1<x_i(1)<x_i(2)<\cdots <x_i(n_i)\le L_{i-1}+n_i \\[1mm]
 (i=1,2,\ldots,\ell) \end{array} \right. \right\}.
\label{3.xset}
\end{equation}
We define the state $f_X$ recursively as
\begin{align*}
 f_j &:= \underbrace{I(\emptyset)\cdots I(\emptyset)}_{P_{\ell-j+1}-P_{\ell-j+2}-1}I(X_j)f_{j-1}
     \quad (\,j=1,2,\ldots,\ell\,),\\
 f_X &:= f_\ell
\end{align*}
where $X_j=\big\{ x_j(k) \big\}_{k=1}^{n_j}\subset X \in {\cal X}_Y$. 
Note that, from the definition of an $10$-insertion, 
$I(\emptyset)$ is the procedure needed to insert $\fbox{\hskip-1.2mm 10\hskip-1.1mm}\,$s between $10$: 
\begin{eqnarray*}
 f &= & 0011100111000001101000111000000, \\
 I(\emptyset)f &= & 00111\fbox{\hskip-1.2mm 10\hskip-1.1mm}0011
 1\fbox{\hskip-1.2mm 10\hskip-1.1mm}0000011
 \fbox{\hskip-1.2mm 10\hskip-1.1mm}01\fbox{\hskip-1.2mm 10\hskip-1.1mm}000
 111\fbox{\hskip-1.2mm 10\hskip-1.1mm}000000
\end{eqnarray*}
and $f_X \in \Omega_Y$ by construction.
We also define $\tilde{\Omega}_Y$ by
\[
\tilde{\Omega}_Y:=\left\{\,f_X\,\big|\,X\in{\cal X}_Y\,\right\}. 
\]

\begin{Lemma}
\label{Lemma2}
\begin{equation}
\langle s_1,s_2,\ldots,s_N \rangle_Y
=\frac{1}{L|\tilde{\Omega}_Y|}
\sum_{f\in\tilde{\Omega}_Y}\sum_{k=1}^L f(k+s_1)f(k+s_2)\cdots f(k+s_N). 
\label{3.lemeq1}
\end{equation}
\end{Lemma}
\Proof
By virtue of the definition of $f_X$, $\tilde{\Omega}_Y$ is the set of states 
with conserved quantities $Y$ and the last entry of the $10$ sequence 
is one of the $0$s that are not marked in the time evolution rule, {\it i.e.},
$\DIS f_X(L)=(T_L f_X)(L)=0$.
By defining the shift operator $S$ by $(Sf)(n):=f(n+1)$, 
and $(S^kf):=S(S^{k-1}f)$ $(k=1,2,\ldots)$ with $S^0f:=f$ and for sets 
\[
S^k\tilde{\Omega}_Y:=\left\{\,S^kf_X\,\big|\,X\in{\cal X}_Y\,\right\}\qquad (k=1,2,\ldots,L),
\]
we find 
\[
^\forall f\in\Omega_Y,\quad \sharp \Big\{\,k\,\Big|\,f\in S^k\tilde{\Omega}_Y\ (k=1,2,\ldots,L)\,\Big\} =L_0. 
\]
Note that $S^Lf=f$. Since $|\Omega_Y|=\frac{L}{L_0}|\tilde{\Omega}_Y|$, 
\begin{align*}
 \langle s_1,s_2,\ldots,s_N \rangle_Y
 &= \frac{1}{|\Omega_Y|}\sum_{f \in \Omega_Y} f(s_1)f(s_2)\cdots f(s_N) \\
 &= \frac{1}{L|\tilde{\Omega}_Y|} \sum_{k=1}^L \sum_{f\in S^k\tilde{\Omega}_Y} f(s_1)f(s_2)\cdots f(s_N) \\
 &= \frac{1}{L|\tilde{\Omega}_Y|} \sum_{k=1}^L \sum_{f\in\tilde{\Omega}_Y} f(s_1+k)f(s_2+k)\cdots f(s_N+k). 
\end{align*}
Thus we obtain \eqref{3.lemeq1}.
\qed

\begin{Proposition}
\label{Propositiofform}
For $X \in {\cal X}_Y$, $f_X$ is explicitly given as
\begin{equation*}
 f_X(n)=u_n^0(X)
\label{3.fandun}
\end{equation*}
where
\begin{eqnarray}
 u_n^t(X) &:= & \eta_{n+1}^{t-1}(X)-\eta_{n+1}^t(X)-\eta_n^{t-1}(X)+\eta_n^t(X), 
 \nonumber \\[2mm]
 \eta_n^t(X) &:= & \max_{{m_{ij}\in\Z,}\atop {i\in[\ell];\ j\in[n_i]}} 
 \left[\ \sum_{i=1}^\ell \sum_{k=1}^{n_i} m_{ik}\big( tP_i-n-x_i(k) + L+k+1+\frac{Z_i}{2}\big) \right. 
 \nonumber \\
 && \hskip3cm \left. -\sum_{i=1}^\ell \sum_{k=1}^{n_i} \sum_{j=1}^\ell \sum_{h=1}^{n_j} 
 m_{ik}\Xi_{ikjh}m_{jh}\ \right], 
 \label{xeta} \\[2mm]
 \Xi_{ikjh} &:= & \frac{Z_i}{2}\delta_{ij} \delta_{kh} +P_{\max[i,j]}, \nonumber \\
 Z_i &:= & L-2\left( P_i\sum_{j=1}^i n_j +\sum_{j=i+1}^\ell n_jP_j \right). \nonumber 
\end{eqnarray}
\end{Proposition}
\Proof
From Proposition~\ref{prop:MIT2008}, $f_X$ is determined by the parameters $W_n$ and $a_n$ $(n=1,2,\ldots,s)$.
Here $W_n$ is the amplitude of the $n$th soliton and $a_n$ is its position, 
{\it i.e.} the position of the $n$th $10$-wall, counting from the right. 
From the definition of the position and of the amplitude of a soliton, 
it follows that both can be determined from $10$ insertions. 
Because of the way $f_X$ was constructed, the set $\DIS \{x_j(k)\}_{k=1}^{n_j}$ corresponds 
to the position of $n_j$ solitons with amplitude $P_j$, 
though it does not directly gives their position.
Hereafter we shall refer to a soliton with amplitude $P$ as a $P$-soliton.
By considering the relation between the position of a soliton and $10$-insertions, 
we find that the position of the $k$th $P_j$-soliton counting from the right is $L-x_j^{(\ell)}(k)+2$, 
where $x_j^{(\ell)}(k)$ is determined recursively: 
we define $x_j^{(i)}(k)\enskip \big(i\in[\,\ell\,],\ j\in[\,i\,],\ k\in[n_j]\big)$ as 
\begin{align*}
 x_j^{(i)}(k)
 &:= x_j(k) +(P_j-P_{i+1})(2\beta_j(k)+2k-1) \\
 & \qquad \quad +\sum_{s=j+1}^i 2(P_s-P_{i+1})\alpha_j^{(s)}(k) -k+1
\end{align*}
where 
\begin{align*}
 \alpha_j^{(i)}(k) &:= \sharp\big\{\,r\in[n_i]\,\big|\,L_{i-1}+n_i-x_i(r)+1>g_j^{(i)}(k)\,\big\}, \\
 \beta_1(k) &:= 0,\qquad 
 \beta_i(k) :=\sum_{s=1}^{i-1} \sharp \Big\{\,r\in[n_s]\,\Big|\,
               g_s^{(i)}(r) >L_{i-1}+n_i-x_i(k)+1\,\Big\}, \\[2mm]
 g_j^{(i)}(k) &:= \max \left\{\,m\in[L_{i-1}+n_i]\,\left|\,
  \begin{array}{l} m=L_{i-1}-x_j^{(i-1)}(k)+1 \\[1mm]
   \qquad +\sharp\big\{\,r\in[n_i]\,\big|\,L_{i-1}+n_i-x_i(r)+1<m\,\big\} 
  \end{array} \right. \right\}.
\end{align*}
Note that $x_j^{(i)}(1)<x_j^{(i)}(2)<\cdots<x_j^{(i)}(n_j)$. 

Recalling the fact that $\sharp\big\{\,r\in[d]\,\big|\,L+d-j_r+1<g_k'\,\big\}$ in \eqref{gk} 
is the number of inserted $\underline{10}$s, on the left of 
the $k$th soliton (here we do not count the inserted $\underline{10}$s as solitons),
the concrete meaning of these variables becomes clear: 
$\alpha_j^{(i)}(k)$ denotes the number 
of $P_i$-solitons on the right of the $k$th $P_j$-soliton, 
and $\beta_j(k)$ denotes the number of solitons 
with amplitudes less than $P_j$, to the right of the $k$th $P_j$-soliton. 

Since $\DIS \{L-x_j^{(\ell)}(k)+2\}_{j=1, k=1}^{\ell, \hskip4mm n_j}$ is the complete set 
of positions of the solitons, there exists a one to one mapping 
$\rho:\big\{\,(j,k)\,\big|\,j\in[\,\ell\,],\,k\in[n_j]\,\big\} \to [\,s\,]$ such that 
\[
 a_{\rho(j,k)} =L-x_j^{(\ell)}(k)+2. 
\]

From these recursion relations we have
\begin{align*}
 x_j^{(\ell)}(k)
 &= x_j(k) +P_j\big(2\beta_j(k)+2k-1\big) +\sum_{i=j+1}^\ell 2P_i\alpha_j^{(i)}(k) -k+1 \\
 &= x_j(k) +2\left\{ P_j\big(\beta_j(k)+(k-1)\big) 
    +\sum_{i=j+1}^\ell P_i\alpha_j^{(i)}(k) \right\} +P_j-k+1.
\end{align*}
Since the position of the $k$th $P_j$-soliton is $a_{\rho(j,k)}$, $W_{\rho(j,k)}=P_j$ and the set of amplitudes 
of the solitons on the right of the $k$th $P_j$-soliton is nothing but $\DIS \big\{W_h\big\}_{h=1}^{\rho(j,k)-1}$.
From the definition of $\alpha_j^{(i)}(k),\ \beta_j(k)$,
\begin{align*}
\alpha_j^{(i)}(k)&=\sharp \left\{W \in \big\{W_h\big\}_{h=1}^{\rho(j,k)-1}\, \Big| \, W=P_i \right\}, \\
\beta_j(k)&=\sharp \left\{W \in \big\{W_h\big\}_{h=1}^{\rho(j,k)-1}\, \Big| \, W> P_j\right\}
\end{align*}
and
\[
\sharp \left\{W \in \big\{W_h\big\}_{h=1}^{\rho(j,k)-1}\, \Big| \, W=P_j\right\}=k-1. 
\]
Thus we obtain 
\[
x_j^{(\ell)}(k)= x_j(k) +\sum_{h=1}^{\rho(j,k)-1} 2\min\big\{W_{\rho(j,k)},W_h\big\} +W_{\rho(j,k)}-k+1. 
\]
Therefore we find a concrete expression of $a_{\rho(j,k)}$, and \eqref{xeta} is immediately obtained 
from \eqref{aeta} and \eqref{aetab}. 
\qed

From Lemma~\ref{Lemma2} and Proposition~\ref{Propositiofform}, we immediately obtain the following theorem:
\begin{Theorem}
\label{cygene}
Let ${\cal X}$ be the set defined in \eqref{3.xset} we have
\begin{equation*}
 C_Y(d_1,d_2,\ldots,d_{N-1}) 
 =\frac{1}{L|{\cal X}|} \sum_{X\in {\cal X}} \sum_{n=1}^L u_n(X) \prod_{i=1}^{N-1} u_{n+d_i}(X), 
\end{equation*}
for $u_n(X)\equiv u_n^0(X)$ as given in \eqref{xeta}.
\end{Theorem}

\section{Concluding remarks}
\label{sec:summ}
In this article, we investigated correlation functions for the PBBS 
and obtained explicit forms for $1$-point and $2$-point functions 
at short distances.
We also give expressions 
in terms of ultradiscrete theta functions for general $N$-point functions.
Investigating their asymptotic properties and to clarify the relation to 
correlation functions for quantum integrable systems are problems that will be addressed 
in the future.

Finally we should comments on the time averages of quantities in the PBBS.
The time average:
$$C_f(d_1,d_2,\ldots,d_{N-1})=\frac{1}{L|{\cal T}_f|} \sum_{t=1}^{{\cal T}_f} 
\sum_{n=1}^L (T_L^tf)(n) \prod_{j=1}^{N-1} (T_L^tf)(n+d_j)$$
where ${\cal T}_f$ is the fundamental cycle of $f\in\Omega_L$ 
depends not only on the conserved quantities of the state 
but, in general, also on the initial state $f$ itself. 
For example, the conserved quantities of the states $f_1=0100100$ 
and $f_2=0101000$ are the same, but
$C_{f_1}(3)=\frac{1}{7}$ and $C_{f_2}(3)=0$. 
Hence, in general, $C_f(d_1,d_2,\ldots,d_{N-1}) \ne C_Y(d_1,d_2,\ldots,d_{N-1})$ 
even for $f \in \Omega_Y$.
Note that, for the $1$-point function $C_f(\emptyset)$, we can easily show that
$${}^\forall f \in \Omega_Y, \quad C_f(\emptyset)=C_Y(\emptyset)=\frac{M}{L}.$$

\section*{Acknowledgement}
The authors wish to thank Ralph Willox for useful comments.

\appendix\section{Example of values for the correlation function}
\label{sec:app2}
From Theorem~\ref{cygene}, we obtain the following examples. 
\begin{description}
\item[(a)] $L=12;\ P_1=3,\ n_1=1;\ P_2=1,\ n_2=2:$
 \begin{equation*}
  C_Y(\emptyset)=\frac{5}{12},\quad 
  C_Y(1)=\frac{1}{6},\quad C_Y(2)=\frac{13}{84},\quad 
  C_Y(3)=\frac{19}{126},\quad C_Y(1,2)=\frac{5}{84}; 
 \end{equation*}
\item[(b)] $L=14;\ P_1=2,\ n_1=2;\ P_2=1,\ n_2=2:$
 \begin{equation*}
  C_Y(\emptyset)=\frac{3}{7},\quad 
  C_Y(1)=\frac{1}{7},\quad C_Y(2)=\frac{5}{49},\quad 
  C_Y(3)=\frac{82}{441},\quad C_Y(1,2)=0; 
 \end{equation*}
\item[(c)] $L=14;\ P_1=3,\ n_1=1;\ P_2=1,\ n_2=3:$
 \begin{equation*}
  C_Y(\emptyset)=\frac{3}{7},\quad 
  C_Y(1)=\frac{1}{7},\quad C_Y(2)=\frac{5}{28},\quad 
  C_Y(3)=\frac{69}{392},\quad C_Y(1,2)=\frac{5}{112}; 
 \end{equation*}
\item[(d)] $L=14;\ P_1=3,\ n_1=1;\ P_2=2,\ n_2=1;\ P_2=1,\ n_2=1:$
 \begin{equation*}
  C_Y(\emptyset)=\frac{3}{7},\quad 
  C_Y(1)=\frac{3}{14},\quad C_Y(2)=\frac{3}{28},\quad 
  C_Y(3)=\frac{13}{112},\quad C_Y(1,2)=\frac{1}{16}. 
 \end{equation*}
\end{description}


\end{document}